\documentclass{jltp}

\title{Coherence between superconducting edge states in
superconducting periodic arrays of artificial defects}
\author{C.C. Abilio$^1$, L.  Amico$^{2,4}$, Rosario Fazio$^{2,3}$,
and B. Pannetier$^1$}
\address{$^1$CNRS-CRTBT, Laboratoire associ\'e
\`a l'Universit\'e Joseph Fourier,\\
25 Av.  des Martyrs, 38042 Grenoble Cedex 9, France\\
$^2$Dip. di Metodologie Fisiche e Chimiche (DMFCI),
Universit\`a di Catania,\\
viale A. Doria, I95219 Catania, Italy\\
$^3$Istituto Nazionale di Fisica della Materia (INFM),
Unit\`a di Catania\\
$^4$Dpto de Fisica Teorica de la Materia Condensada,\\
Universidad Autonoma de Madrid, Madrid E-28049, Spain. }

\input epsf
\epsfverbosetrue

\begin{document}

\maketitle
\begin{abstract}

The transition line of superconducting arrays of holes
exhibits a rich field structure due to the interference of
superconducting states nucleated at the holes edges.
We studied by means of resistance measurements their
effect on the $T^{*}_{c}(H)$ line as a function of transverse magnetic
field using regular arrays of nanofabricated micron size holes.
The arrays transition fields are higher than for the bulk. Moreover we found a 
nontrivial field modulation of the $T^{*}_{c}(H)$
line with an inversion, with increasing field, of the modulation
concavity which we assigned to a crossover from a collective to an 
isolated edge state regime.
The high field regime is well described by the
nucleation at a single hole in an infinite film.
The modulation at low fields was found to be dominated by the
interference of neighbor edge states when the inter-hole distance $w$
becomes comparable to the coherence length $\xi(T^{*}_{c})$.
A comparison between arrays of different hole shape shows the
influence of geometry on the type of interaction established, which
can described either as a superconducting wire network
or as a weak link array.

PACS numbers: 74.76,74.60,74.25
\end{abstract}

\bigskip

\section{Introduction}

The effect of artificial pinning centers on the vortex dynamics of
type-II superconductors has attracted a great interest in both
fields of high $T_{c}$ and conventional low $T_{c}$ superconductors.
Low $T_{c}$ materials are particularly interesting due to
the development of nanofabrication techniques which
enable the introduction in a controlled way of nanofabricated
defects.
These patterned samples constitute relatively simple systems for the
study of how geometric parameters such as defect size, shape and/or 
defect density influence vortex 
dynamics\cite{Bezryadin,Fiory,Moschalkov}.

Previous experiments on superconducting arrays of holes have
shown that the transition line 
is dominated by surface superconducting states nucleated around 
the hole boundaries~\cite{Bezryadin}. As a result, the array 
transition occurs at fields $H^{*}_{c3}$ higher than the nucleation 
fields in the bulk $H_{c2}$, and the transition line $T^{*}_{c}(H)$ 
exhibits a non-trivial field modulation
due to flux quantization of the edge states over the two geometric 
lengths of the problem: the hole surface and the array elementary cell.
In particular, it was identified a crossover with decreasing field 
from an isolated hole regime to a low field regime dominated by the interaction between edge 
states.
Though the behavior at high fields was well understood, the nature of 
the interactions at low fields remained unclear.

In this paper we address in detail the collective regime at low 
fields when the distance between adjacent holes edges $w$ is 
comparable to the array lattice constant $a$.
The relevant parameters are the external field, the temperature 
dependent coherence length
$\xi(T^{*}_{c})$, the lattice constant $a$, the hole size,
and the inter-hole distance $w$.
We will try to put forward the role of these parameters on the
crossover to the collective regime and the
type of interaction established at low fields.
The two extreme cases, the isolated hole in an infinite film\cite{Buzdin} and the 
superconducting wire 
network~\cite{Alexander,Rammal_PhysB} (vanishing hole separation) are quite well known.

In section \ref{sec:Exp} we present transport measurements on a
superconducting square array where the aspect ratio hole
size/lattice parameter is approximately two. These results
are compared in section \ref{sec:Discussion} with our previous work~\cite{Bezryadin}
on a different array with similar aspect hole/lattice parameter but
with twice the distance between hole edges.
The distinct behavior found at low fields will be discussed as arising from
distinct nucleation processes, determined by the ratio
$w/\xi$.
In section \ref{sec:WNT}  we analyze the case where  $w/\xi(T^{*}_{c})$ is lower than the
critical ratio $1.84$~\cite{Saint-James}. Nucleation
is then dominated by thin wire superconducting edge states and the 
coupling between them is well described by the formalism of
superconducting wire networks~\cite{Alexander,Rammal_PhysB}.
In section \ref{sec:WLA} we analyze the case $w/\xi(T^{*}_{c})>1.84$ where edge nucleation
is determined at the single hole edge
but $w$ is still small enough to allow a weak overlap between
neighbor edge state wave functions.
In this regime the low field behavior can be described using a
model based on a weak link interaction between
adjacent edge states. This simple model is able to capture the
essential features of the crossover previously discovered~\cite{Bezryadin}.

\section{Experimental Details}
\label{sec:Exp}
The array sample consists of a thin film of aluminum ($80$~nm)
patterned with a regular square array of nanofabricated holes.
The lattice spacing is $4.0~\mu$m and holes have a square shape
with $1.85\pm 0.01~\mu$m side length, the distance between neighbor
hole edges being $w=2.15\pm 0.01~\mu$m as determined by SEM microscopy.
We shall refer to this sample as sample
A.  The full array size is $1\times 1$~cm$^{2}$ corresponding to a total of
$6.25 \times 10^{6}$ holes.

The patterning was defined on a monolayer PMMA coated 2" Si wafer by Deep UV
photolithography using a high precision chromium mask\cite{chromium}.  The sample is then prepared
using conventional $\textit{lift-off}$ techniques after thermal evaporation
of pure
aluminum over the resist mask on a UHV chamber.  An homogeneous thin
film of aluminum evaporated at the same time and submitted to the
same fabrication steps is also measured for a reference of the
patterning process effect on the material parameters.

Both samples were studied by conventional
four-probe resistance
measurements using an AC four terminal
resistance bridge at
a $33$~Hz frequency and a measuring current
of $2$~nA.
Assuming an uniform current distribution
over the array, the current density per wire is
$4.5\times 10^{-4}$~Acm$^{-2}$.
We used a similar in-line geometry for the voltage contacts (spiral-shaped) in the 
array sample as in our previous work~\cite{Bezryadin}, placed at a distance 
of $2.8~$mm
from each other in the array center to avoid
short circuits from the sample borders. 
For the reference sample we simply attached gold 
(non-superconducting) wires by ultrasonic bounding.
No overshoot in the $R(T)$ curves
was observed, in contrast to the previously studied
reference sample~\cite{Bezryadin}
where the spiral contacts were used and small cusps in the R(T) 
curves were present.
\footnote{We remark that the R(T) curves for array A
in non zero field do not exhibit the 
double transition observed on array B, which was then assigned to 
a first transition at $T_{c2}$ and to a second transition at higher 
temperatures $T^{*}_{c}$. We now believe the observation of the $T_{c2}$ 
transition in array B was possible due to the transition of the 
20~$\mu$m strips between patterned fields (of about 
$300\times 300~\mu$m$^2$). In the case of array A, which is an homogeneous 
pattern over a 1~cm$^{2}$ surface, the double transition is not observed.}

Resistance measurements
as a function of temperature were also performed
at several magnetic fields between 0 and 5~mT.
From the zero field R(T) measurement we estimated several material
parameters. Using a two-dimensional Aslamazov-Larkin fit~\cite{Aslamazov},
we estimate a BCS transition temperature
$T_{co}=1.262$~K and a normal state resistance of $R_{n}=0.099~\Omega$.
The resistance per single wire of length $a=4.0~\mu$m and width
$w=2.15~\mu$m is then $r_{n}=0.354~\Omega$ and the normal state
resistivity $\rho_{n}=1.49~\mu \Omega$cm.
Using $v_{F}=2.03\times 10^{8}$~cm/s we obtain an electronic mean
free path $l_{el}=26.5$~nm. The zero field transition temperature defined
at half of the normal state resistance are 1.263~K and 1.265~K for the
patterned sample and the homogeneous thin film, respectively, the
transition width being 3~mK for both samples.
We summarize on Table 1 some parameters of array A and the parameters of
the sample previously studied~\cite{Bezryadin}, array B.
This array consists on an aluminum thin film ($80$~nm thickness)
patterned with a square array of circular holes of diameter $2r=4.26
~\mu$m and a lattice parameter of $9.0~\mu$m.

The field dependence of the nucleation temperatures $T^{*}_{c}(H)$
is determined using a heating feedback technique that keeps the sample
resistance at a constant value while the magnetic induction field is
smoothly varied by small increments of $0.2~\mu$T.
This method enables
us to attain a fine field tuning of the array transition line.
Several resistance criteria between 0.01 and $0.7 R_{n}$ were used.
The nucleation temperatures using the sweeping field method
agree by less than 1~mK with those obtained for the same
resistance criteria from R(T) measurements,
indicating a good temperature regulation
attained by the feedback method.

\begin{figure}
\epsfxsize=12 cm
\epsffile{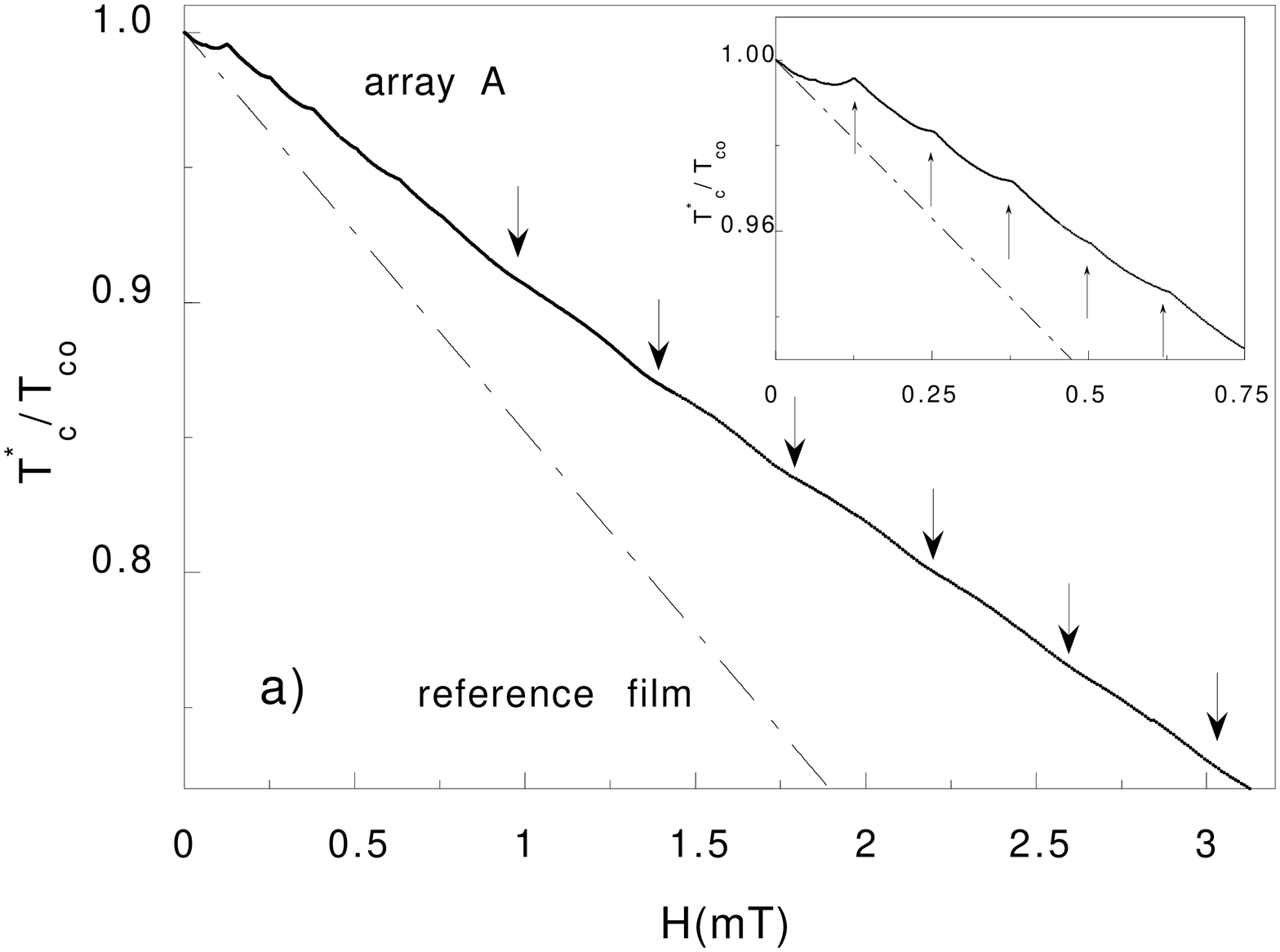}
\epsfxsize=10.7 cm
\epsffile{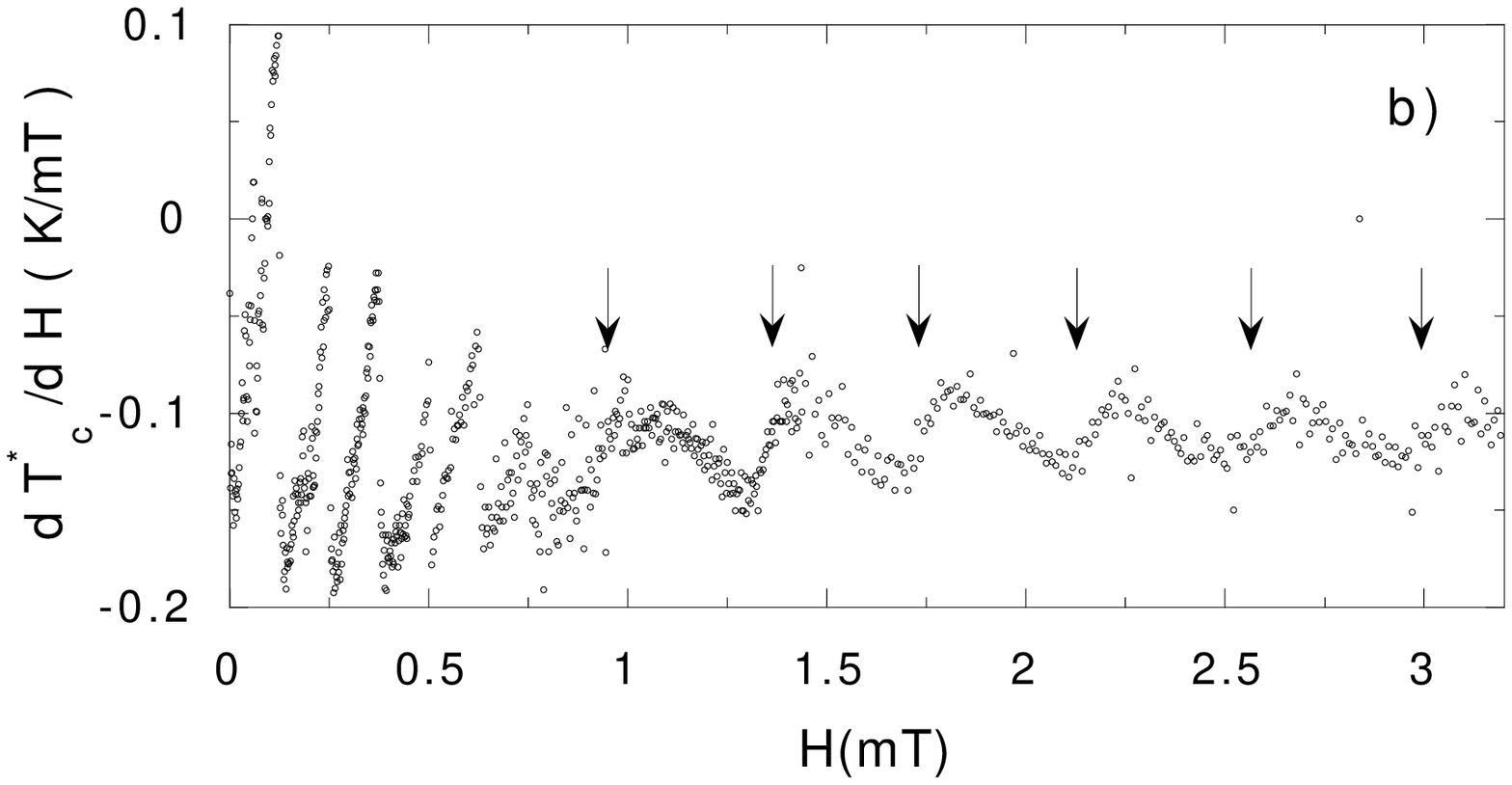}
\caption{a) Field dependence of the
	superconducting transition line $T^{*}_{c}(H)$
	of array A (solid line) and of the reference sample
	$T_{c2}(H)$ (dashed line).
	Two types of field modulation are clearly identified for array A:
	downward,
	large period oscillations with dips at half integers of $\Phi_{o}$
	per hole (down arrows) and upward oscillations of shorter
	period with cusps at integers of $\Phi_{o}$ per array elementary
	cell (inset:up arrows);
	b) $T^{*}_{c}(H)$ slope for array A as a function of H.
	The change of magnetic period due to the crossover from collective to single hole
	regime is quite visible.}
\label{Tc(H)}
\end{figure}

The transition line $T^{*}_{c}(H)$ of array A for criteria $0.4 R_{n}$ 
is displayed in Fig.~\ref{Tc(H)}.a) as a function of applied field.  
For comparison it is also displayed the reconstructed bulk transition 
line, $H_{c2}(T^{*}/T_{co})= 
\left(1-T^{*}/T_{co}\right)\Phi_{o}/2\pi\xi^{2}(0)$, using the 
reference film coherence length at $T=0$~K, $\xi(0)=220$~nm 
(determined from the initial linear slope of the reference film 
transition line).  The array transition line is clearly above the 
estimation for the bulk.  Besides, the non-trivial field modulation 
associated to the single hole and collective regimes are clearly 
identified.  At fields below $0.75$~mT it is characterized by periodic 
upward cusps, with a magnetic period of exactly $H=0.128$~mT, that 
correspond to a quantum flux enclosed on a square cell of side length 
$a=4.0~\mu$m, indicating the presence of phase coherence over the 
array lattice.  At higher fields, the transition line exhibits 
downward dips occurring with a larger magnetic period.  These large 
period oscillations correspond to the single hole regime discussed 
previously.  In this case, the magnetic period is not constant since 
it depends on the effective surface formed by the hole radius and the 
surface superconducting sheath surrounding the hole.  The average 
magnetic period is $0.543$~mT, which corresponds to an effective 
square surface of side length $l_{hole}=1.94~\mu$m, comparable to the 
array hole size of $1.85~\mu$m.

These field modulations were found
for all the criteria used in the
transition line measurement.
The crossover between the two field
regimes is well illustrated on Fig.~\ref{Tc(H)}.b) where the sawtooth
variation of the transition line slope $dT^{*}_{c}(H)/dH$, at low
field gives place to a smoother variation at higher fields, along 
with the change of magnetic period.

\section{Discussion}
\label{sec:Discussion}

\subsection{Extraction of the energies for nucleation of
superconductivity}

The role of the
nucleation processes involved will be discussed in terms of
the energies for nucleation of superconductivity at a given field,
$\epsilon_{nucl}$.
The energy $\epsilon_{nucl}$ can be obtained by finding the
lowest eigenvalue solution of the linear Ginzburg
Landau differential equation
\begin{equation}
\frac{\hbar^{2}}{4m}\left[\frac{\nabla}{i}-\frac{2e}{c}\textbf{A}\right]^{2}
\psi(\textbf{r})=\epsilon_{nucl}\psi(\textbf{r})
\end{equation}
which fulfills the given boundary conditions on the order parameter
$\psi(\textbf{r})$.

This approach is valid when we can neglect spatial variations of
$|\psi(\textbf{r})|$, such as thin films or wires of thickness $\ll \xi$
or when the applied magnetic field reduces $|\psi(\textbf{r})|$ to a value much
smaller than the equilibrium  amplitude $|\psi_{\infty}|$  achieved
deep inside the bulk superconductor. The regime of validity is
then usually restricted to temperatures close to $T_{co}$.

An alternate approach is to extract $\epsilon_{nucl}$ at a given field from
the measured $T^{*}_{c}(H)$ using the 
relation~\cite{Abrikosov,deGennes},
       \begin{equation}
           ln \left[  \frac{T^{*}_{c}}{T_{co}}  \right]=
           \Psi_{D}\left[  \frac{1}{2}   \right]
           -\Psi_{D}\left[ \frac{1}{2}+\frac{\epsilon_{nucl}}{4\pi
           k_{B}T^{*}_{c}}\right ]
           \label{lnT}
       \end{equation}
where $\Psi_{D}(x)=\Gamma'(x)/\Gamma(x)$ is the digamma
function.
This relation describes the depression of $T^{*}_{c}(H)$ relative to
$T_{co}$ due to a magnetic perturbation.
Though it was initially established by Abrikosov and
Gor'kov for magnetic impurities~\cite{Abrikosov}, it can be generalized to
all pair breaking perturbations which destroy the time reversal
symmetry of Cooper pairs~\cite{deGennes,Maki}, if
the scattering time of the electron pair is short enough for 
their relative phases be randomized
by the perturbation.

It can thus be applied to a dirty superconductor in strong external
magnetic fields (and only surrounded by insulators), if the mean free path
$l_{el}$ is much smaller than all
sample dimensions and $\xi_{o}$ or in the case of a small superconducting
particle with all dimensions $\ll \xi_{o}$.
The sample thickness must be smaller than $\xi$ and the penetration
depth to ensure the penetration of the magnetic field.
At temperatures close to $T^{*}_{c}(H)$, 
$\epsilon_{nucl}$ is the energy required to break the Cooper pair
thus destroying superconductivity.
When $T^{*}_{c}=0$, (or $H=H^{*}_{c3}(0) )$, $\epsilon_{nucl}$ coincides with the
BCS superconducting gap $1.76 k_{B}T_{co}$.
Close to $T_{co}$ the digamma function can be expanded around 1/2
and the $T^{*}_{c}(H)$ depression is linear in $\epsilon_{nucl}$,
$T_{co}-T^{*}_{c}(H)=\epsilon_{nucl} \pi/8 k_{B}$.

The advantage of using Eq.~\ref{lnT} on the determination of
$\epsilon_{nucl}$ is that it remains valid down to all temperatures 
and in strong magnetic fields.
The linear Ginzburg-Landau
results can be recovered if the temperature
dependent coherence length is defined as
$\xi^{2}(T)=D\hbar/\epsilon_{nucl}$, where $D=1/3v_{F}l_{el}$
is the coefficient for electronic diffusion.
For sample A, $D=180$~cm$^{2}$s$^{-1}$ obtained from
the mean free path $l_{el}=26.5$~nm. Using $\xi^{2}(0)=\hbar
D/1.76k_{B}T_{co}$ we estimate $\xi(0)=250$~nm.

\begin{figure}
\epsfxsize=12 cm
\centerline{\epsffile{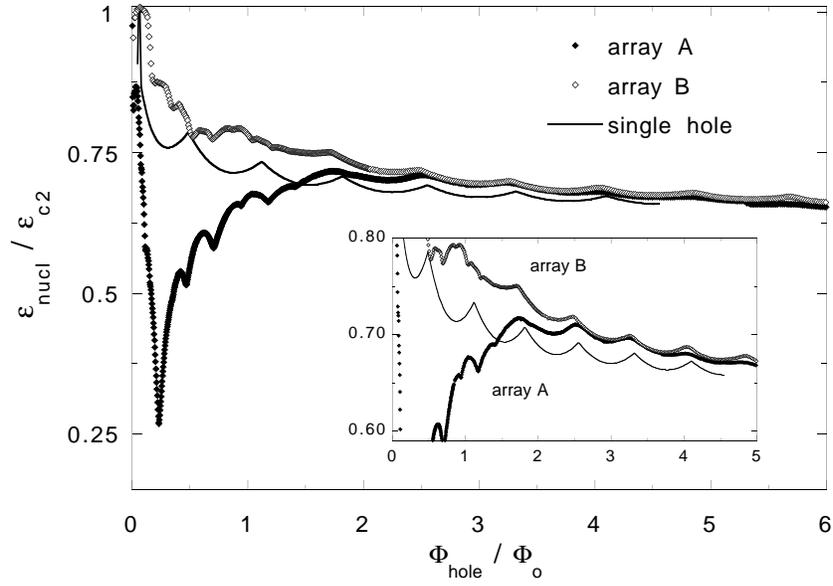}}
\caption{Normalized nucleation energies $\epsilon_{nucl}/\epsilon_{c2}$
	as a function of magnetic field (in units
	$HS_{hole}/\Phi_{o}=\Phi_{hole}/\Phi_{o}$), for sample A (solid
dots), sample B (open diamonds) and the theoretical calculation for a
	cylindric cavity in an infinite thin film (solid line).}
\label{En/Ec2}
\end{figure}

We calculated $\epsilon_{nucl}(H)$ for arrays A and B using
Eq.~\ref{lnT} and the experimental $T^{*}_{c}$ at the given applied field H.
These results are displayed in Fig.~\ref{En/Ec2}
after being normalized by
the nucleation energy on the bulk, $\epsilon_{c2}=hD
H/\Phi_{o}$, with the
same coherence length as the array.
For comparison, we also represent
the theoretical $\epsilon_{nucl}/\epsilon_{c2}$ for a circular hole on an infinite film
(solid line)~\cite{Bezryadin-Ovch}.
In fact, the representation $\epsilon_{nucl}/\epsilon_{c2}$ is equivalent
to the inverse ratio of the nucleation
fields $H_{c2}/H^{*}_{c3}$, that close to $T_{co}$ acquires the
Ginzburg Landau form $H_{c2}/H^{*}_{c3}=(1-T^{*}_{c}/T_{co})~H_{c2}(0)/H$, 
using $H_{c2}(0)=\Phi_{o}/2\pi\xi^{2}(0)$ and the coherence length
as defined above.

From Fig.~\ref{En/Ec2} it is clear that in the high field regime both
samples are very well described by the theoretical single hole case.
With decreasing fields both arrays deviate from the single hole 
description, with the appearance of the collective field modulation, 
periodic on $\Phi_{o}$ per array cell, and with upward concavity.
It is in this regime that the samples present a strikingly distinct 
behavior.
The reduced energies $\epsilon_{nucl}/\epsilon_{c2}$
for sample B and for the single hole tend to an overall increase with
decreasing field, reaching 1 at zero field.
In contrast, for sample A,
$\epsilon_{nucl}/\epsilon_{c2}$ decrease with decreasing field, dropping 
well below the single hole line.
This means that in this regime, array A presents a ratio 
$H^{*}_{c3}/H_{c2}$ higher than the classical limit of 1.69 for an 
infinite surface sheath\cite{Saint-James}.
Comparing the array geometric parameters, they both have similar aspect ratios 
$w/a$, $0.54$~(array A) and $0.53$~(array B), respectively
and similar ratios of the superconducting volume over the array cell 
volume $V_{s}/V_{cell}$, $0.79$~(array A) and $0.82$~(array B).
We thus believe the distinct low field behavior is associated to 
the different ratio $w/\xi(T^{*}_{c})$, which close to $T_{co}$ 
controls the process of edge nucleation and the type of coupling.
In fact, both samples loose the collective behavior 
for $w/\xi(T^{*}_{c})>3$.

On the following subsections we shall analyze these features
by describing the array nucleation energy as coming from two main
contributions: the nucleation energy of
the single edge state and the coupling energy
between neighbor edge states.

\subsection{Wire networks of wide strands}
\label{sec:WNT}

In this section we focus on the low field behavior of array A.
In this regime, the array energy is a sum of
the nucleation energy on a individual wire in parallel field $\epsilon_{strip
\parallel}$ and a coupling energy, that can be described within the framework of
superconducting wire networks theory~\cite{Alexander,Rammal_PhysB}.

The case of superconducting wire networks of narrow wires is well
understood. It can be treated as a periodic array of
superconducting islands strongly coupled to each other
by thin superconducting wires of length $a$ ($\xi>>a$) and width $w<<\xi$.
Neglecting superconducting fluctuations, the coupling energy can be
computed within mean-field theory by solving the linearized
Ginzburg-Landau equations at each node of the network.
For a regular square lattice (same length of all strands) the order
parameter
$\psi_{i}$ on each island $i$ will be coupled to the first neighbor
sites through field dependent phase factors as,

     \begin{equation}
         4~\cos{(u)}~\psi_{i}=\sum_{j}^{}\psi_{j}\exp(-i\gamma_{ij})
         \label{tb}
     \end{equation}
where 4 is the lattice coordination,
$\gamma_{ij}=2\pi/\Phi_{o}\int_i^j {\bf A} \cdot d {\bf l}$ is the phase
factor along
the wire linking site $i$ to a site $j$, $\bf A$ the vector potential,
and $u=a/\xi(T)$ is the strand length in units of $\xi(T)$. The coupling energy 
is then given by,  $\epsilon_{wnt}=\hbar Du^{2}/a^{2}$.

Eq.~\ref{tb} is equivalent to a tight binding equation in a potential 
with the same geometry and tight binding energy $\epsilon_{tb}=4~
\cos u$. The network coupling energy $\epsilon_{wnt}=\hbar Du^{2}/a^{2}$ 
can then be 
expressed in terms of the tight binding ground state level $\epsilon_{tb}$
with $u=\arccos{(\epsilon_{tb}/4)}$~\cite{Hofstadter,Rammal_PRL}.
However, this relation is only valid in the limit $w>>\xi$.
Taking into account the finite thickness of the wires, a more complex 
relation between $\epsilon_{tb}$ and $u$ is obtained,
$\epsilon_{tb}=4~\cos u + 4~\tan(u w/2a)~\sin u - u w/a$~\cite{Wang}.
This result was established for zero external field but it can still
be applied at low fields while there are no vortices
in the wires.

\begin{figure}
\epsfxsize=12cm
\centerline{\epsffile{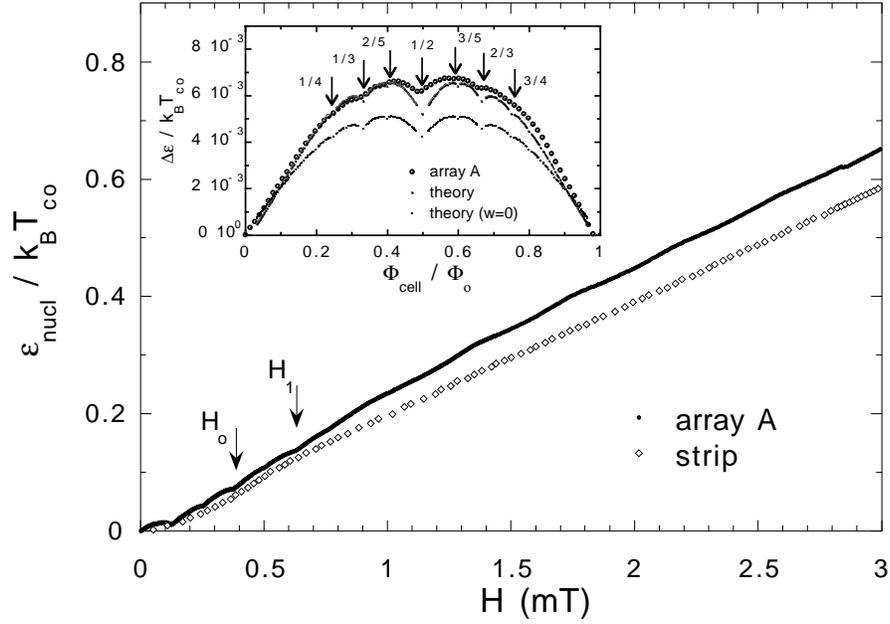}}
\caption{Field dependence of the nucleation energies
	of array A (solid dots) and for a strip of
	width $w=2.15\mu$m (open diamonds), normalized by $k_{B}T_{co}$.
	In the field range $H_{o}<H<H_{1}$ interstitial vortices appear within strands.
	Inset: Coupling energy $\epsilon_{wnt}$ for array A (solid dots) and the 
	theoretical $\epsilon_{wnt}$ for a superconducting wire network with 
	$w=0$ (small dots; lowest curve) and taking into account the wire thickness (small dots,
	upper curve) as a function of reduced field $\Phi_{cell}/\Phi_{o}$ between 0 and 1.
	The main dips position at rationals $p/q$ are indicated by down arrows.}
\label{energy-ech-vs-slab}
\end{figure}

In the inset of Fig.~\ref{energy-ech-vs-slab}
is displayed the coupling energy of array A and the theoretical $\epsilon_{wnt}$ energy as a function of
magnetic flux per array elementary cell $\Phi_{cell}$,
for $\Phi_{cell}/\Phi_{o}$ between 0 and 1.
The coupling energy for array A is obtained from
$\epsilon_{nucl}$ (calculated using Eq.~\ref{lnT} and the experimental $T^{*}_{c}(H)$)
after subtracting the parabolic energy contribution due to 
edge nucleation in the wires, $\epsilon_{strip~\parallel}$.
The theoretical $\epsilon_{wnt}$ is obtained from the ground state 
tight binding energy calculated for
rational values of $\Phi_{cell}/\Phi_{o}=p/q$, with $q<30$ and $p<q$.
A very good agreement is obtained between the experimental and the theoretical 
data which take into account the
wire thickness. The theoretical results in the limit $w=0$ 
lead to a smaller energy.
Since we are close to $T_{co}$, the variation of $T_{co}-T^{*}_{c}(H)$ due
to the coupling (after subtracting the linear dependence on H) corresponds
to $\epsilon_{wnt}\pi/8k_{B}$.

Besides the fundamental dips at
$\Phi_{cell}/\Phi_{o}=0,1$ and
at 1/2, additional dips can
be identified at rational values $\Phi_{cell}/\Phi_{o}=p/q$, for
$q=3,4$ and 5.
This field structure is a
manifestation of the 
interference of quantum states over cells of size $qa\times 
qa$~\cite{Rammal_PRL}.
With increasing field, the fine field structure
becomes less pronounced and only the fundamental dips remain until
$\Phi_{cell}/\Phi_{o}=8$.
At higher fields the single hole regime is
recovered.

In fact, the crossover to the single hole regime is associated to a 
crossover from a two-boundary to a one-boundary nucleation process in 
the network wires.
This explains the smaller values of
$\epsilon_{nucl}/\epsilon_{c2}$ when compared to the 
classical limit $\epsilon_{nucl}/\epsilon_{c2}=0.59$ for nucleation in 
an infinite surface sheath\cite{Saint-James} or for nucleation in a single hole.
Such as for a thin slab in a parallel field, edge nucleation in the array
at low fields is controlled by two boundary 
conditions imposed at the edges of adjacent holes.

The field dependence of $\epsilon_{strip~\parallel}$ for a strip or
slab of intermediate thickness $w$, is strongly
dependent on $w/\xi(T^{*}_{c})$.
Below a critical thickness $w<1.84~\xi(T^{*}_{c})$
(thin wire limit)
nucleation starts symmetrically at both surfaces and the
maximum of the order parameter occurs at middle distance between
them.
In this limit
$\epsilon_{strip~\parallel}= H^{2}w^{2}(\pi hD/6 \Phi_{o}^{2})$.
With increasing field, when $w>1.84~\xi(T^{*}_{c})$,
$\epsilon_{strip~\parallel}$ deviates from
the parabolic field dependence as the order parameter solutions
at each surface pull apart, their sobreposition giving rise to
nodes along the middle plane of the wire and equidistant of
$\Delta L \approx \Phi_{o}/Hw(1-1.84\xi)$.
With increasing field the vortex pattern becomes more complex, until
the interference between the surface solutions become negligible
(compared to $k_{B}T^{*}_{c}$) and the one-boundary solution is recovered.
In this limit $\epsilon_{strip~\parallel}$ approaches the surface sheath 
result $0.59 hDH/\Phi_{o}$.

\begin{figure}
\epsfxsize=12cm
\epsffile{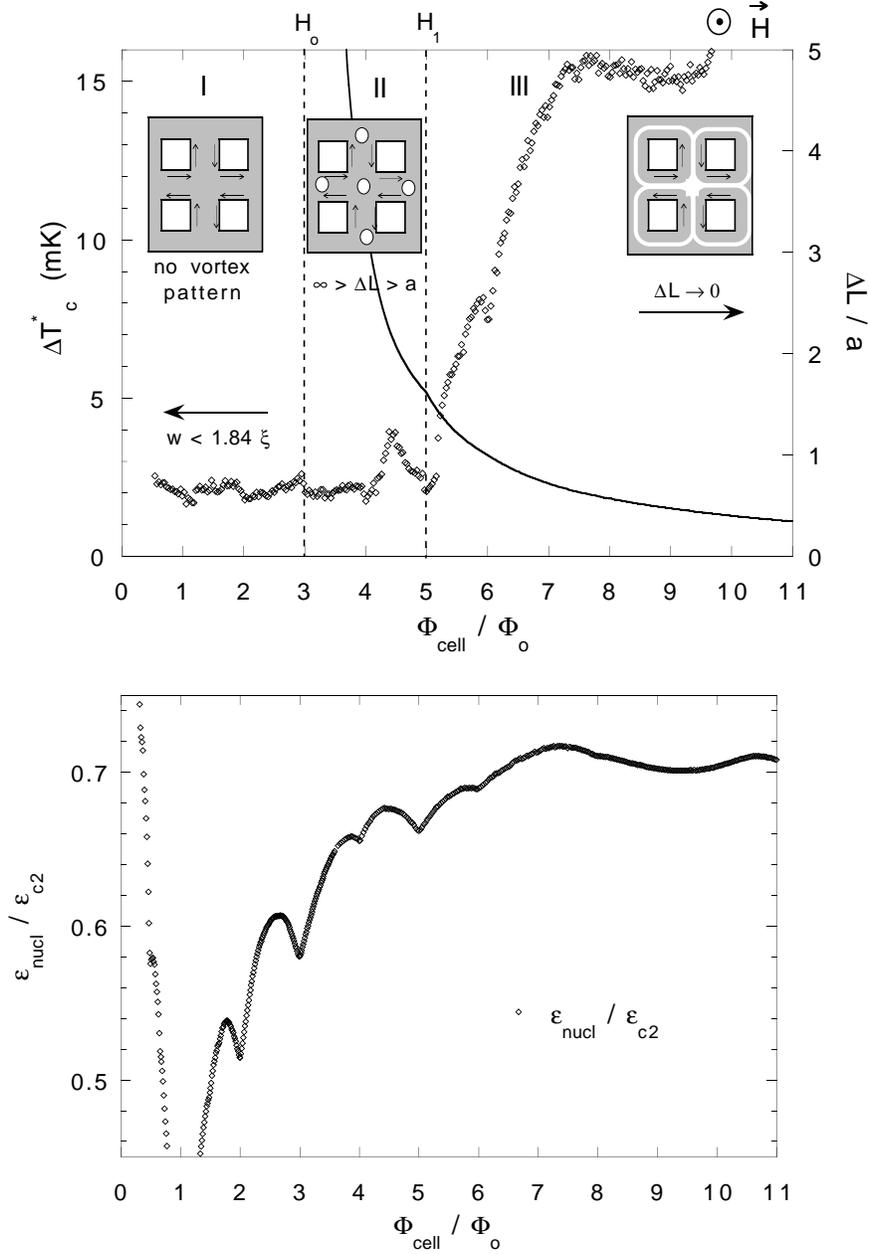}
\caption{a) Resistive transition width $\Delta T^{*}_{c}$ of array A (open
diamonds) as a function of the reduced flux $\Phi_{cell}/\Phi_{o}$,
and comparison with the normalized distance $\Delta L/a$
between interstitial vortices for a thin wire (solid line).
An oversimplified picture of the vortex patterns developed
within the wires is represented. Three main regions can be
identified: (\textrm{I}) $w<1.84\xi(T)$, nucleation starts
symmetrically and there is no vortex in the
wires; (\textrm{II}) $w>1.84\xi(T)$ and $\infty >\Delta L/a\leq 1$,
nodes of the order parameter appear at interstices due to
the interference of neighbor edge wave functions (white dots);
(\textrm{III}) $\Delta L/a<1$ and decreases with
increasing field until the surface solutions
become independent and the single edge states are localized around
each hole. b) field variation of the nucleation energy
$\epsilon_{nucl}/\epsilon_{c2}$ for array A.}
\label{dT}
\end{figure}

All these features were discussed previously~\cite{Saint-James,Fink,Schultens}.
Here we are interested in comparing the envelope of the energy curve
for array A and the nucleation energy for a  strip, $\epsilon_{strip~\parallel}$.
On Fig.~\ref{energy-ech-vs-slab} is displayed $\epsilon_{nucl}$ for array A and 
$\epsilon_{strip~\parallel}$
for a wire with the same width $w$ as the array strands as a function of 
applied field.
We can identify several similitudes. On the single hole regime the
main dependence of $\epsilon_{nucl}$ is linear on H such as the field
dependence of $\epsilon_{strip~\parallel}$ in the surface sheath limit, in
agreement with the dominance of one-boundary nucleation at the edges of each individual hole.
With decreasing field both curves deviate from the linear field dependence due to the
emergence of
interference between adjacent surfaces solutions. This deviation occurs
near the field $H_{1}=0.65$~mT which corresponds to the position of the
first important collective dip of $\epsilon_{nucl}/\epsilon_{c2}$ at $\Phi_{cell}/\Phi_{o}=5$ (see also
Fig.~\ref{dT}.b).
Below the field $H_{o}\approx 2.75~\Phi_{o}/\pi 
w^{2}$ ($0.39$~mT) nucleation starts
symmetrically\cite{Saint-James} in the strip and the order parameter is 
maximum at middle distance between adjacent edges.
The occurrence of symmetric nucleation on a low field regime
explains why the wire network description is still valid
for arrays of wide strands.
In the regime $H<H_{o}$, the collective dips of
$\epsilon_{nucl}$ approach the parabolic envelope
$\epsilon_{strip~\parallel}$, since at integers values of
$\Phi_{cell}/\Phi_{o}$ the costs in coupling energy are minimum.
These results thus indicate that the small values of $\epsilon_{nucl}/\epsilon_{c2}$
for array A at low fields 
are simply related with the two-boundary nucleation process.

On the other hand, since the crossover from
two-boundary to one-boundary nucleation is associated to the appearance of
interstitial nodes of the order parameter within strands, we expect a
broadening of the array resistive transition with increasing field due to
these weakly bounded interstitial vortices.
In Fig.~\ref{dT} is represented the field variation of the resistive
transition width $\Delta T^{*}_{c}(H)$, for array A obtained by
subtracting the
transition lines $T^{*}_{c}(H)$ measured for resistive criteria
$0.7 R_{n}$ and $0.03 R_{n}$. For comparison, we also represent the
field variation of the distance $\Delta L$ between nodes on
a single wire, normalized by the array lattice parameter $a=4~\mu$m.

At fields $H\leq H_{o}=0.39$~mT (region \textrm{I},
$\Phi_{cell}/\Phi_{o}\leq 3$)
the transition width at integers $\Phi_{cell}/\Phi_{o}$ is
the same as in zero field since there are no order parameter nodes
in the strands, only coreless vortices inside holes.
At integers $\Phi_{cell}/\Phi_{o}$, the flux quanta per array 
cell corresponds to the quanta enclosed by each hole.
The transition is then sharpened since
every hole encloses the same number 
of flux quanta.
At intermediate
values of $\Phi_{cell}/\Phi_{o}$ phase fluctuations lead to a
small broadening of the transition (of about 1~mK).

At fields $H_{o}<H<H_{1}$ (region \textrm{II},
$3<\Phi_{cell}/\Phi_{o}<5$), the first
nodes of the order parameter
are expected to appear within the strands with a separation $\Delta L$
that drops from infinity to values comparable to the
array lattice parameter.
In this regime there will be a 
competition between increasing the flux per hole by $\Phi_{o}$ or
follow the increase of field
by accommodating vortices at interstitial positions in the array wires.
The increase of the transition width for $\Phi_{cell}/\Phi_{o}$ 
between 4 and 5 illustrates the presence of a few loosely bound 
vortices.
This observation is in agreement
with Fig.~\ref{dT}.b) where the field dependence of
$\epsilon_{nucl}/\epsilon_{c2}$
represented as a function of $\Phi_{cell}/\Phi_{o}$ shows that
$\epsilon_{nucl}/\epsilon_{c2}$ at $\Phi_{cell}/\Phi_{o}=4$ is higher
than the adjacent dips.
At $\Phi_{cell}/\Phi_{o}=5$ every hole encloses $4~\Phi_{o}$ which favors the decrease of
$\epsilon_{nucl}/\epsilon_{c2}$. Also, the distance between vortices 
in the strands should be of the order of $a$ and, in that case, they  
can occupy positions at the array interstices forming a stable
sub-lattice, which decreases the resistive transition width.

In region \textrm{III}, the distance between interstitial vortices 
should drop below $a$ and the transition width broadens
considerably due to these weakly bounded  vortices.
The distance between nodes is further reduced with increasing
field until edge states at each hole become independent.

We can see from Fig.~\ref{dT}.b) that at $\Phi_{cell}=8~\Phi_{o}$,
the field modulation of $\epsilon_{nucl}/\epsilon_{c2}$ inverts concavity
once the single hole nucleation becomes dominant.
If for $\Phi_{cell}/\Phi_{o}$ between 5 and 8, the 
additional flux occupies interstitial positions, the flux 
per hole remaining equal to $4~\Phi_{o}$, $\epsilon_{nucl}/\epsilon_{c2}$ should 
meet the single hole curve at
its $4^{th}$ oscillation, in agreement with the results
presented on Fig.~\ref{En/Ec2}.

\subsection{Weak link array of edge states}
\label{sec:WLA}

We do not expect the description of a wire network of wide strands to hold if
holes edges are pulled apart and/or shaped to circles.
This is the case of sample B where the minimum inter-hole
distance is $w=4.74~\mu$m ( $\sim $ twice compared to
array A). The field for symmetric
nucleation is $H_{o}\approx 0.08$~mT for a stripe with the same thickness.
However since the distance
between hole edges is not constant (varying from 4.74 to $9~\mu$m
around the hole perimeter) $H_{o}$ should be further reduced. As a 
consequence 
the parabolic envelope of $\epsilon_{nucl}$ is not observed.
The coupling between neighbor edge states is still present though.
Fundamental dips of $\epsilon_{nucl}$ at integers values of $\Phi_{cell}/\Phi_{o}\leq 7$ are
clearly observed although we find no fine field structure~\cite{footnote1}.
At low field $\epsilon_{nucl}/\epsilon_{c2}$ is
lower than the single hole calculation indicating that the energy
contribution due to the overlap between edge states is important.

In the next sections we present a simple model which is able to capture the
transition from the single hole to the collective behavior for array B.
The application of this model is based on the assumption that in a
narrow temperature region $T_{c2}<T<T^{*}_{c}$ the superconducting
edge states are localized close to the hole boundaries and the system
behaves like an array of weak links.
The fast oscillations observed at low fields are then related to the
collective behavior of the array. Since there is
nonzero overlap between the wave functions of different edges,
the energy should be higher than that of a single hole.
One consequence is that the nucleation field $H^{*}_{c3}$
is lower than that of a single hole,
as it is observed in the experiments. The single
hole behavior is naturally recovered at higher fields where
the overlap between neighboring edges vanishes.

A Ginzburg-Landau approach is suitable to study the problem. For the
single hole, an analytical solution is available due to the cylindric
symmetry of the problem~\cite{Bezryadin2}.
A variational approach for the
determination of the order parameter has been analyzed by Buzdin~\cite{Buzdin}.
The agreement with the exact
solution is good except at low fields where the variational approach
predicts a region with zero flux through the hole which is
not present in the exact solution. Buzdin's variational  approach,
however, has the great advantage of being applicable also in the many
holes case where an analytical solution is impossible to obtain.

We proceed as follows. We first describe an improved variational ansatz
for the single hole case by using a trial order
parameter of three parameters.
We then construct an effective Ginzburg-Landau free energy which
takes into account, in an approximate way, the edge coupling and we determine
the resulting nucleation energy.

\subsection{Variational approach : single hole }

In this section we introduce the new variational wave function for
the evaluation of the nucleation energy in a superconducting film containing a
hole. Following Buzdin~\cite{Buzdin}, the Ginzburg-Landau free energy
( ${\cal F }_N$ is the free energy of the normal state) is given by

       \begin{equation}
           {\cal F }  - {\cal F }_N= {{\hbar^2}\over{4 m}}\int  d^2r \left [
           \left |\left (-i \nabla -{{2 \pi}\over{\Phi _o}}
           {\bf A}\right )\psi (\vec{r})\right |^2 -
           \frac{1}{\xi^2} \left |\psi (\vec{r}) \right |^2
           \right ] \;\;\;
           \label{fhole}
       \end{equation}

The fourth order term has be ignored since we are interested in the
phase boundary. The surface critical field can be determined with a
variational procedure by determining the minimum of the functional
       \begin{equation}
           {\cal F }  - {\cal F }_N= 0
       \end{equation}
within the class of trial wave functions $\psi (\vec{r})$ which
satisfy the proper boundary condition at the
superconductor-insulator interface
       \begin{equation}
           \left [{{\partial \psi}\over{\partial \rho}}\right ]_{\rho=R}=0
       \end{equation}
where $R$ is the radius of the hole.  For a hole containing $m$ flux
quantum, the order parameter has the following form in cylindric
coordinates $(\rho, \, \phi, \, z)$,
$$\psi=\frac{1}{\sqrt{2 \pi}} F(\rho) e^{i m \phi }$$.
It is possible to improve the result obtained by ~\cite{trial} by using a
three-parameters trial order parameter of the form
       \begin{equation}
           F(x)=\left [1+\alpha (x-x_0)^{\eta}+
           \beta (x-x_0)^{\zeta} \right ]
           \exp{\left [-{{\gamma}\over{2}} (x-x_0)^2\right ]}
           \label{trial3}
       \end{equation}
where $\alpha, \beta ,\gamma$ are the variational parameters and  the
dimensionless
quantities $x =  \rho (e H/c \hbar)^{1/2}$,
$x_0= R(e H/c \hbar)^{1/2}=(\Phi_{hole}/\Phi_o)^{1/2}$ has been introduced.

\begin{figure}
\epsfxsize=10cm
\centerline{\epsffile{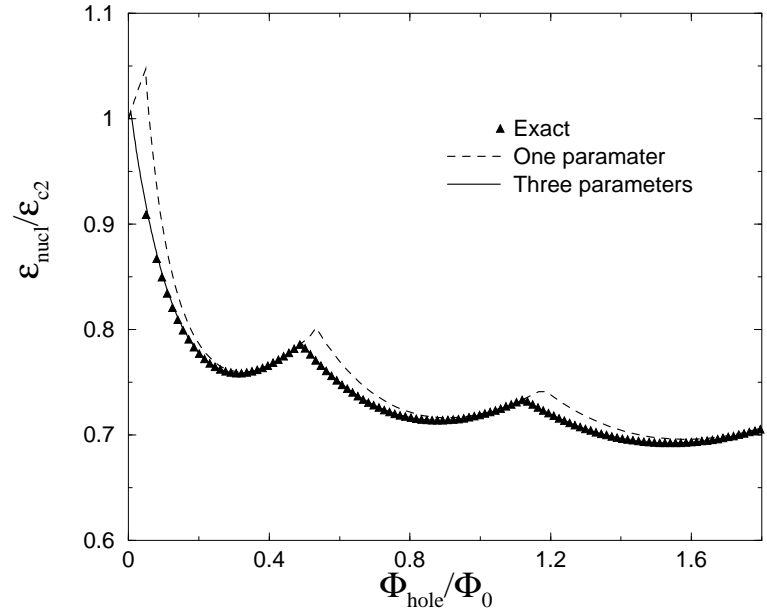}}
\caption{Variational approximations for
$\epsilon_{nucl}/\epsilon_{c2}$.
The thick curve is obtained using the three parameter variational wave
function.
For comparison the curve obtained from the one parameter variational 
function is reported (dashed line).}
\label{singlevar}
\end{figure}

For the case of a single hole an exact solution for the nucleation energy
is expressed in terms of the eigenvalues of the Kummer's
equation~\cite{Bezryadin2,benoist}.
We compare our results (the best results have been obtained for
$\eta=2$  and $\zeta=2.05$) with the exact calculation and with the
one parameter trial function~\cite{Buzdin}. This is shown in
Fig.~\ref{singlevar}.
The trial function $F(x)$ given by Eq.~(\ref{trial3}) has a maximum at
$(x - x_0)^2 \sim (2-\gamma / (\alpha +\beta))/\gamma$, a feature which
is also present in the exact solution. This seems to improve considerably
the accuracy of the variational approach~\cite{tinkham}.

The result obtained by using Eq.~(\ref{trial3}) improves considerably
the agreement with the exact solution. In particular the spurious $m=0$
is shrunk to very low fields ($< 0.03~\Phi_o$).

\subsection{Variational approach : array of holes}

\begin{figure}
\epsfxsize=7cm
\centerline{\epsffile{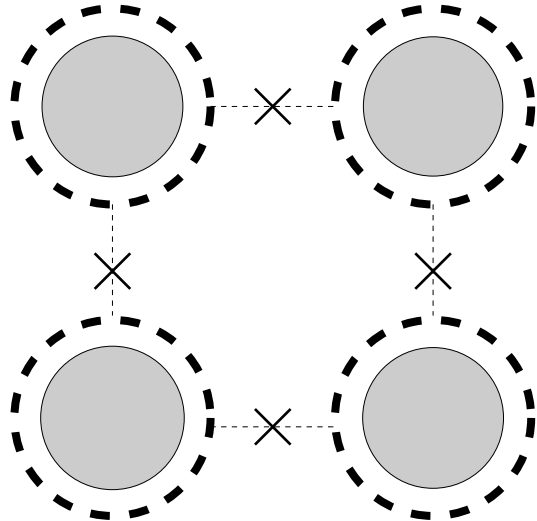}}
\caption{Elementary cell of a square array of holes (grey circles).
The edge states localized around each hole (thick dashed lines) are 
weakly coupled to first neighbor edge states. The coupling is 
indicated with crosses.}
\label{model}
\end{figure}

We obtain an approximate expression of the nucleation energy for a
regular array  using the variational wave function introduced in the
previous section
and an appropriate Ginzburg-Landau free energy which takes into account
the overlap between edge states of adjacent holes. The model is
sketched in Fig.~\ref{model}, we will consider an array of edges connected
by weak links (dashed  lines in the figure). The crossover between
the single hole and the collective behavior stems from the interplay of
flux quantization on each hole and frustration effects through the
elementary array cell.

The GL free energy which we propose has the following form
       \begin{equation}
           {\cal F }  - {\cal F }_N
           =
           \sum_i {\cal F }_i +
           \sum_{\langle i,j \rangle} {\cal F }_{ij}
       \end{equation}
where ${\cal F }_i$ is the free energy defined in
Eq.~(\ref{fhole})
and the subscript $i$ identifies the hole in the array.
${\cal F }_{ij}$ takes into account the overlap between the
holes.

We choose a ${\cal F }_{ij}$ of the following form
       \begin{equation}
           {\cal F}_{ij} = {{\hbar^2 }\over{4 m a^2}}
           \int d^2 r \left[ \alpha_{o}
           \left | \psi_i(\vec{r}) -\psi_j (\vec{r}) \right |^2 +
           (\alpha_1+ \alpha_2\delta_{ij} )
           \left | \psi_i(\vec{r}) \left |\right|\psi_j (\vec{r}) \right |
           \right]
           \label{fint}
       \end{equation}
The effect of the external magnetic field will be included in
Eq.~\ref{fint}
by means of a Peierls substitution. We assume that the overlap
is restricted to a small area (see the dashed lines in
Fig.~\ref{model}) and
that the hole array can be treated as a weak link array.
The first term in Eq.~\ref{fint} is modified as follows
       \begin{equation}
           \left | \psi_i -\psi_j \right |^2
           \rightarrow
           |\psi_i|^2 + |\psi_i|^2 -
           2|\psi_i ||\psi_j| \cos(\phi_i-\phi_j -\pi m -A_{ij})
       \end{equation}
where
$
A_{i,j}=(2 \pi)/(\Phi_o) \int_i^j {\bf A} \cdot d {\bf l}
$
is calculated along the path indicated by the dashed lines in
Fig.~\ref{model} (we assumed that all the holes contain the same
number $m$ of flux quantum).

The terms in Eq.~\ref{fint} can be understood considering that
the order parameter in the array of holes can be given
as $\psi = \sum_i \psi_i$. Substituting this expression in the GL free
energy of Eq.~\ref{fhole}, one generates the various terms given
above. In general since the approach is phenomenological, the various
contributions enter with different coefficients ($\alpha_{o}, \alpha_1,\alpha_2$).
It is important to stress that the total wave function does {\em not}
satisfy the proper boundary conditions around the holes because of the
exponential tails of the edge states of the neighboring island.
Therefore this approach breaks down at very small field when the overlap
becomes too strong.

The agreement of the experiments with the single hole result in the
high flux regime indicates that phase fluctuations do not drive
the phase transition. In this case we can approximate the phase
dependent part of the free energy by its ground state energy
$$
\epsilon (\Phi_{cell}/\Phi_o) = \frac{2}{zN}\sum_{\langle i,j \rangle}
    \langle \cos(\phi_i-\phi_j -\pi m -A_{ij})\rangle _{GS}
$$
where $N$ is the number of holes in the array, $z$ is the coordination
number, $\Phi_{cell}/\Phi_o$ is the magnetic flux per elementary cell 
of the array and $\langle \dots \rangle _{GS}$ means the ground state
configuration of the phases $\phi_i$.

Going over the same steps as Buzdin~\cite{Buzdin} the nucleation energy
for an array of holes is obtained by minimizing the following
functional over the trial function introduced in Eq.~\ref{trial3}
       \begin{eqnarray}
           {\epsilon_{nucl}\over{\epsilon_{c2}}} & = &
           {\cal I}
           \int_{ x_0}^\infty \frac {d x}{x}
           \left[ \,(m-x^2)^2 F^2(x)
           +x^2 [F^{'}(x)]^2  +  g
           \frac {R^2}{a^2}
           \frac {1}{4x_0^2}x^2
           [F(x)]^2
           \right] \nonumber \\
           & + & g_1\frac {R^2}{a^2}\frac {1}{4x_0^2}
           \left[2-g_2 \; \epsilon(\Phi_{cell}/\Phi_o)\right]
           \langle F(x_i) F (x_{j})\rangle
       \end{eqnarray}
where the overlap integral
       \begin{equation}
           \langle F(x_{i}) F (x_{j})
           \rangle ={\cal I}
           \int  d \vec{x}
           F(\vec{x} ) F ( \vec{x} + \frac{\vec{w}}{x_0})
       \end{equation}
the normalization
$$
{\cal I}^{-1} = 2 \int_{x_0}^\infty dx \; x F^2(x)
$$
has been introduced. The new parameters $g,g_1,g_2$ are easily expressed
as a function of the $\alpha$'s.

\begin{figure}
\epsfxsize=10cm
\centerline{\epsffile{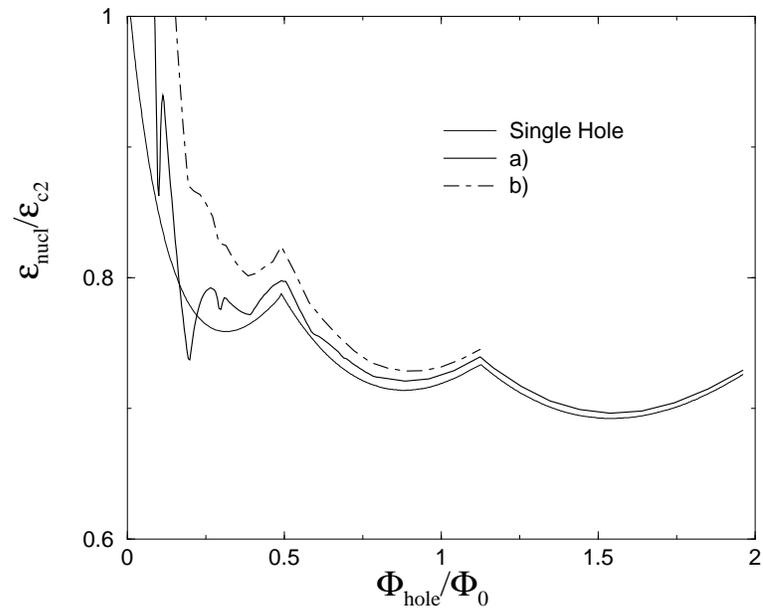}}
\caption{The normalized nucleation energy
$\epsilon_{nucl}/\epsilon_{c2}$ for a regular array of holes obtained by
minimizing the functional as defined in the text
for various parameter values: a) g=0.1 g1=0.4 g2=1.6;
b) g=0.2 g1=0.7 g2=0.9.}
\label{arrayvar}
\end{figure}

In Fig.~\ref{arrayvar}
we present the results obtained for various
values of the three parameters $g$,$g_1$ and $g_2$.
We can observe that the presence of the Josephson coupling increases 
the array nucleation energy, relatively to the isolated hole case, and 
introduces cusps of different concavity superimposed over the main 
single hole background at low flux when the 
coupling becomes stronger. These features were found in the 
experiment of Bezryadin \textit{et al} ~\cite{Bezryadin}.
At very low fields $\epsilon_{nucl}/\epsilon_{c2}$ rapidly increases, 
as also
seen in the experiments. It may be explained by the fact the the wave
functions extends over various lattice constants and therefore the
system does not show any surface superconductivity. However this model
breaks down at flux typically of the order of $0.15~\Phi_{hole}$.

A word of caution is needed at this point. Although the main features of
the crossover are found one should be aware that the model is still
too simplified to aim at a quantitative comparison with the experiments.
In particular the choice of the values of the constants $g$ is not related
to any microscopic model which allow to justify the numerical parameters.
It is, however, rewarding that most of the qualitative features are captured
by our model. A further step might be to apply the techniques developed 
by Palacios \textit{et al}~\cite{palacios} to the present problem.

\section{Conclusion}

In conclusion, we tried to show how geometric parameters such as
the inter-hole and the array lattice constant can influence the
transition temperature of periodic superconducting arrays
in a magnetic field.

We found a clear crossover with decreasing field from a single
to a coupled edge state regime for both samples.
The behavior of $\epsilon_{nucl}/\epsilon_{c2}$ in the single edge 
regime is very similar and the arrays transition depend exclusively on the
magnetic flux per hole
area.
However, we remark that a quantitative comparison of the energy ratios
$\epsilon_{nucl}/\epsilon_{c2}$ between samples need some caution since
they are extremely dependent on the
estimation of $\xi(0)$ and small errors can lead to a shift
of the $\epsilon_{nucl}/\epsilon_{c2}$ values.

The low field behavior of the studied samples is
representative of two distinct coupled 
systems: the superconducting wire network (strong coupling) and the weak 
link array. In both cases the coupling is mediated by the frustration 
induced by the applied field over the array elementary cell, 
however the type of coupling depends on the inter-hole distance $w$,
the defect shape and the coherence length at $T^{*}_{c}(H)$ which controls the
edge nucleation process.
In the case of the weak link array (array B) we have a coupling of
single hole edge states which depend on H through the enclosed flux 
per hole. However, in the low field regime, the enclosed flux does not necessarily 
follow the field increase since the change of the hole winding 
number may be less favorable than placing flux at the array 
interstices where superconductivity is weakened. 
In the wire network case (array A at low fields), we have a strong coupling 
of the order parameter at the array nodes, which is non-zero 
along the wires due to symmetric nucleation. The flux enclosed per hole or per array cell are 
equivalent quantities as long as interstitial vortices are not allowed 
within the array strands. In this regime, the main contribution to the array transition
comes from nucleation in the wires (which
depends only on H).

This explains why the studied samples who have similar geometric 
ratios ($w/a\approx 2$ and $V_{s}/V_{cell}\approx 0.8$) exhibit such a 
different collective behavior.
The relevant parameters are then the inter-hole distance (which 
allows two-boundary nucleation in a wide field regime for sample A)
and the parallel hole edges which favor the thin wire 
nucleation.
For periodic arrays of close enough ($w/\xi(T^{*}_{c})<1.84$) 
and parallel hole edges, the coupling between edge 
states can be described as a wire network coupling and the 
array transition approaches the thin wire 
nucleation at low fields. As a consequence, the 
ratio $H_{c3}/H_{c2}$ ($\epsilon_{nucl}/\epsilon_{c2}$) in these arrays is not
necessarily upper (lower) bounded by the surface sheath result $H_{c3}/H_{c2}=1.69$
($\epsilon_{nucl}/\epsilon_{c2}=0.59$) or the single
hole nucleation limit.

When two-boundary nucleation does not occur, due to a 
higher inter-hole distance and/or due to the hole shape,
the wire network formalism is
not suitable to describe the coupling between edge states,
since the variations of the order parameter amplitude between holes 
cannot be neglected.
In this case the array transition is dominated by 
single hole nucleation and we presented a coupling description, based in a
weak link interaction between single edge states, which is able to recover
the main features of the $T^{*}_{c}(H)$ line on the low field regime: 
the inversion of field modulation concavity and the 
presence of periodic upward cusps.
In addition, it reproduces the experimental observation that the 
array transition occurs at temperatures below the single 
hole nucleation, since the array energy is increased due 
to the overlap between neighbor edge  wave functions.

\section{Acknowledgments}
We would like to thank G. Falci and O. Buisson for useful discussions.
We acknowledge the financial support of the European Community
(Contract FMRX-CT-97-0143).
C.C. Abilio is supported by a grant
PRAXIS XXI (5713/95) of the Portuguese Ministry for Science and
Technology.

\newpage

\newpage

\newpage

\newpage

\newpage

\newpage

\begin{table}
\begin{center}
\begin{tabular}{c|cccccc}
\hline\hline
\\  
$Sample$ &$hole$~$shape$ & $a$~($\mu$m) & $w$~($\mu$m) &
$T_{co}$~[K] & $\xi(0)$~($\mu$m)\\ \\
\hline
\\
$\quad$array A & square &  4.0 & 2.15 &  1.263 & 0.25\\
$\quad$array B & circle & 9.0 & 4.74  &1.25 & 0.25\\
\hline
\end{tabular}
\end{center}
\caption{Some parameters of array A and array B.
The array lattice constant is $a$ and $w$ the
minimum distance between the edges of adjacent holes.}
\end{table}


\begin{thebibliography}{9}
\bibitem{footnote1}
The absence of fine field structure could be interpreted has due to
the presence
of coherence only at a first neighbors level or due to the amplitude
oscillations
being smaller than our temperature resolution.
\bibitem{trial} In Ref.4 Buzdin has chosen a trial form of $F(\rho)$ of the type
$F(x)=\exp{\left [-{{\gamma}\over{2}} (x- x_0)^2\right ]}$.
\end{thebibliography}
\end{document}